\documentclass{article}

\usepackage{amsmath}
\usepackage{amssymb}

\usepackage{PRIMEarxiv}

\usepackage[utf8]{inputenc} 
\usepackage[T1]{fontenc}    
\usepackage{hyperref}       
\usepackage{url}            
\usepackage{booktabs}       
\usepackage{amsfonts}       
\usepackage{nicefrac}       
\usepackage{microtype}      
\usepackage[toc,page]{appendix}
\PassOptionsToPackage{breakable,many}{tcolorbox}
\usepackage{tcolorbox}
\usepackage{xcolor}
\usepackage{lipsum}
\usepackage{fancyhdr}       
\usepackage{graphicx}       
\graphicspath{{media/}}     
\usepackage{enumitem}
\title{Hide and Seek: Fingerprinting Large Language Models With Evolutionary Learning
\thanks{\textit{\underline{Our code}}: 
\texttt{https://github.com/MorpheusAIs/HideNSeek}} 
}

\author{
  Dmitri Iourovitski\\
  \texttt{dmitri.io@utexas.edu}\\
  \and
  Sanat Sharma\\
  \texttt{sanatsharma@utexas.edu}\\
  \and
  Rakshak Talwar\\
  \texttt{rakshaktalwar@gmail.com}
}

\definecolor{goodcolor}{RGB}{0,128,0}
\definecolor{mixedcolor}{RGB}{255,140,0}
\definecolor{poorcolor}{RGB}{220,20,60}
\definecolor{thoughtcolor}{RGB}{70,130,180}

\newenvironment{goodscenario}
  {\begin{tcolorbox}[colback=goodcolor!10,colframe=goodcolor,title=Scenario - LLM Performs Well]}
  {\end{tcolorbox}}

\newenvironment{mixedscenario}
  {\begin{tcolorbox}[colback=mixedcolor!10,colframe=mixedcolor,title=Scenario - LLM Has Mixed Performance]}
  {\end{tcolorbox}}

\newenvironment{poorscenario}
  {\begin{tcolorbox}[colback=poorcolor!10,colframe=poorcolor,title=Scenario - LLM Underperforms]}
  {\end{tcolorbox}}

\newenvironment{agentthoughts}
  {\begin{tcolorbox}[colback=thoughtcolor!10,colframe=thoughtcolor,title=Agent's Thoughts]}
  {\end{tcolorbox}}
  
\definecolor{neutralgray}{RGB}{240,240,240}

\newenvironment{promptcard}
  {\begin{tcolorbox}[
    colback=neutralgray!30,
    colframe=neutralgray!60,
    breakable
  ]}
  {\end{tcolorbox}}

\begin{document}

\maketitle
\begin{abstract}

As content generated by  Large Language Model (LLM) has grown exponentially, the ability to accurately identify and fingerprint such text has become increasingly crucial. In this work, we introduce a novel black-box approach for fingerprinting LLMs, achieving an impressive 72\% accuracy in identifying the correct family of models (Such as Llama, Mistral, Gemma, etc) among a lineup of LLMs. We present an evolutionary strategy that leverages the capabilities of one LLM to discover the most salient features for identifying other LLMs. Our method employs a unique "Hide and Seek" algorithm, where an Auditor LLM generates discriminative prompts, and a Detective LLM analyzes the responses to fingerprint the target models. This approach not only demonstrates the feasibility of LLM-driven model identification but also reveals insights into the semantic manifolds of different LLM families. By iteratively refining prompts through in-context learning, our system uncovers subtle distinctions between model outputs, providing a powerful tool for LLM analysis and verification. This research opens new avenues for understanding LLM behavior and has significant implications for model attribution, security, and the broader field of AI transparency.

\end{abstract}


\section{Introduction}
Hide and Seek is a novel algorithm that uses Large Language Models (LLMs) for the purpose of uncovering the hidden semantic manifold of another LLM, which allows for accurate and robust fingerprinting of the type of family of an LLM. Each LLM within a family has seen relatively similar data and therefore their semantic manifold will be similar to one another. Using the knowledge, we first formulate the semantic manifold hypothesis which provides a theoretical framework for what, if anything, can be utilized as a fingerprint when it comes to a language model. With the framework, we find distinct tells that LLMs have when generating content. \\

\begin{figure}[h]
    \centering
    \includegraphics[width=0.8\textwidth]{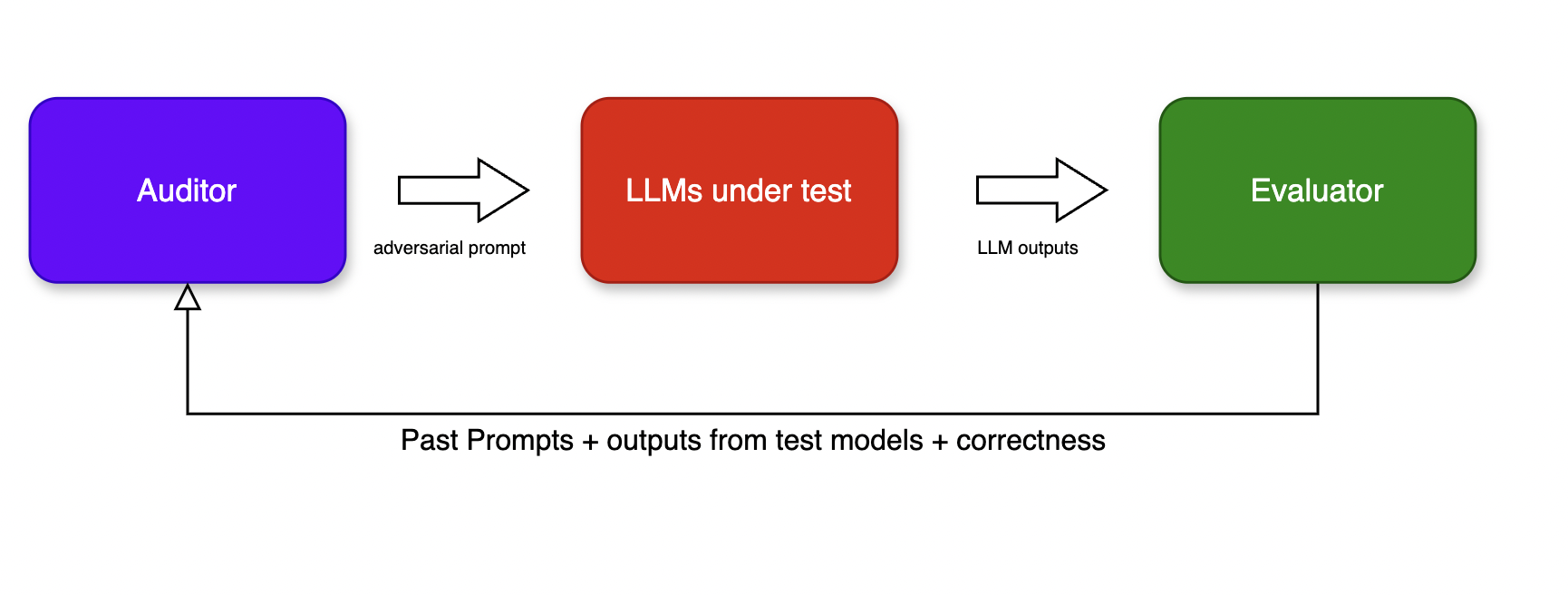}
    \caption{Hide and Seek is an evolutionary algorithm that aims to uncover the hidden semantic manifold of LLMs. The Auditor is tasked with generating "adversarial" prompts that elicit unique responses from the models under test. The responses are then passed to an LLM evaluator which tries to predict the 2 models in the test set that are similar to each other. The model outputs and evaluator correctness is passed back to the auditor to refine the prompt.}
    \label{fig:overview}
\end{figure}

Following a hypothesis for how and why LLM outputs differ, we devise a discriminitive prompt strategy that is aimed at maximizing the diversity of outputs across different families of LLMs. This is done while assuming that all LLM models will be treated as black boxes, with their internal workings and training data remaining unknown. An overview of the approach is shown in Fig \ref{fig:overview}.

Generating a discriminitive prompt is challenging and requires excellent semantic understanding along with a comprehensive understanding of the text. We claim that LLMs are good adversarial prompt generators and can drive the process of discovering the discriminitive prompts. To discover differences and identify related LLMs, we also use an LLM to act as a detective. This idea borrows heavily from previous works that utilize LLMs as evaluators/judges\cite{zheng2023judgingllmasajudgemtbenchchatbot}\cite{liu2023gevalnlgevaluationusing}. The detective explores the outputs of LLMs that are being tested out based on the prompts generated by the auditor who is tasked with coming up with the questions that will maximize the differences amongst outputs.

Our findings can be summarized as follows:

\begin{itemize}
    \item Theoretical formulation of the way in which LLMs generate tokens and how this leads to a fingerprint being present
    \item A methodology that can discover subtle differences between LLMs, and how this enables detection
    \item An evolutionary in-context learning approach that enables LLMs to learn the important distinctions across LLMs
    \item Lays out the challenges and opportunities with fingerprinting various LLMs
\end{itemize}

\section{Semantic Manifold Hypothesis}
The Semantic Manifold Hypothesis (SMH) posits that generative natural language processing (NLP) models, despite their apparent complexity and high-dimensional output space, operate on a significantly lower-dimensional manifold when generating sequences of tokens. This hypothesis extends the traditional manifold hypothesis in machine learning \cite{goodfellow2016deep} to the output space of language models, suggesting that the generative capabilities of these models are more constrained than previously thought.
\subsection{Background}
The manifold hypothesis in machine learning states that real-world high-dimensional data often lies on or near a low-dimensional manifold \cite{fefferman2016testing}. This concept has been crucial in developing dimensionality reduction techniques and understanding the behavior of deep learning models \cite{bengio2013representation}. However, the application of this hypothesis to the output space of generative language models represents a novel perspective.
\subsection{Formulation}
The SMH can be formally stated as follows: Given a sequence of tokens $s = (t_1, t_2, ..., t_n)$, a generative language model $M$ produces a probability distribution over the next token $t_{n+1}$ that lies on or near a manifold $\mathcal{M}_s$ of significantly lower dimension than the full vocabulary space $V$:
\begin{equation}
P_M(t_{n+1} | s) \approx \mathcal{M}_s \subset \mathbb{R}^{|V|}, \quad \dim(\mathcal{M}_s) \ll |V|
\end{equation}
This formulation suggests that the effective dimensionality of the model's output is much smaller than the size of the vocabulary, potentially explaining observed limitations in language model outputs.

\section{Model definition under SMH}
Under the Semantic Manifold Hypothesis, an LLM is really a set of outputs that the specific LLM has the capacity to generate. 

\subsection{Formal Definition}
Let $M_i$ be an arbitrary LLM model, and let $X$ be a specific known model. We define $S_i$ as any sequence of tokens. The probability that $M_i$ is equivalent to $X$ given a sequence $S_i$ is denoted as:

\begin{equation}
    P(M_i = X | S_i)
\end{equation}

We aim to find the sequence $S_x$ that maximizes this probability:

\begin{equation}
    M_x = \underset{S_x}{\operatorname{arg\,max}} \; P(M_i = X | S_x)
\end{equation}

This maximization is achieved when:

\begin{equation}
    M_x \cap M^c = \emptyset
\end{equation}

where $M^c$ represents the complement of $M_x$. This condition implies that the set of tokens that best identifies $X$ shares no overlap with any tokens from the complement of $S_x$.

To achieve this, we seek to uncover $\hat{S}$, a subset of all possible generations of $M$ that is as unique as possible:

\begin{equation}
    \hat{S} \subset \{S : S \text{ is a possible generation of } M\}
\end{equation}

It's important to note that LLMs are tuned to understand prompts, and this is the primary mechanism for interacting with LLMs behind a black box. Therefore, to obtain $\hat{S}$, it is necessary to craft $\hat{P}$, a family of prompts:

\begin{equation}
    \hat{P} = \{P_1, P_2, \ldots, P_n\}
\end{equation}

where each $P_i$ is designed to elicit responses that contribute to the unique identification of the model $X$. 
These prompts are adversarial in nature and aim to extract the following:

\begin{itemize}
    \item Elicit specific responses that expose the model's unique characteristics.
    \item Probe weaknesses or idiosyncrasies in the model's training data or architecture.
    \item Differentiate the model from others by targeting areas where the model's behavior is distinctive.
\end{itemize}

\section{Real World considerations of crafting $\hat{P}$ and $\hat{S}$}

Crafting The Optimal prompts for discovering the optimal Sequence has many real world challenges that must be first addressed in order to proceed with an effective approach using the Semantic Manifold Hypothesis.

\subsection{Intractable Response Exploration}
Uncovering the set of all responses for a generative model is not a tractable problem. Furthermore, many recent works have been focusing on expanding S even more for any given P, such as \cite{zhang2024improvingdiversitycommonsensegeneration} seeks to boost the sets of responses to a given prompt an LLM is capable of outputting. The intersection of the works on diversity and representation in \cite{lahoti2023improvingdiversitydemographicrepresentation} highlights the semantic biases LLMs tend to exhibit when given certain prompts and how to improve representational knowledge within an LLM. These two works illustrate the ever-expanding frontier of possible generations given a prompt.

\subsection{Ambiguity in Feature Space S}
In a real sense, there's a very large ambiguity in which features make S the most salient. On the surface, counting tokens using approaches like n-gram counting \cite{brown1990statistical} or TF-IDF \cite{sparck1972statistical} is unlikely to contain specific enough features ensuring a uniqueness in S. Part of the Semantic Manifold is that an LLM's potential response will use different surface-level tokens that contain the same meaning for a response.

Methods that rely on contextual cues \cite{reimers-2019-sentence-bert, cer-2018-universal} will be misguided by similarly sounding outputs across LLMs, as once again outputs are on a Semantic Boundary defined more by a policy from the Language Model designer \cite{wei2021instructiontuning, rafailov2023direct} than what is found in natural language. Furthermore, both of the above approaches display a length bias where longer texts tend to be predicted as similar to each other due to overlap of tokens as an aggregate.

An intelligent approach that can approximate text features dynamically is a requisite for this kind of problem. 

\subsection{Stochastic Nature of Generative Responses}
Generative Responses exhibit a very large degree of stochasticity and thus are not deterministic. These responses are affected by several features inside the model architecture such as temperature, token confidences (topp) and more.

Recall that $\hat{P}$ represents our family of prompts and $\hat{S}$ is the subset of all possible generations of the model. The many-to-many relationship between prompts and their potential outputs can be represented as:

\begin{equation}
    \hat{P} \xrightarrow{\text{many-to-many}} \hat{S}
\end{equation}

This notation indicates that for any given prompt $p_i \in \hat{P}$, there exists a set of potential outputs $S_i \subset \hat{S}$, where $|S_i| \gg 1$, emphasizing the stochastic nature of the generative process. A good process, as showcased in \cite{dathathri2020plug}, requires careful steering of an LLM towards more unique and constrained responses to minimize the many-to-many relationship as much as possible.

\section{A Game of Hide and Seek - Uncovering the Latent Manifold}

Uncovering the specific model using prompts and outputs is similar to playing a game of hide and seek. In hide and seek, the rules are straightforward: a group of people hides, and a seeker is tasked with discovering their hiding locations. This analogy is fitting because, as studies like \cite{peskin2008hide} showcase, hide and seek requires skills in spatial reasoning, problem-solving, and introspection to uncover potential hiding spots. Similarly, in the context of identifying a model, the "seeker" (or auditor) must use prompts to elicit unique responses from the model, effectively "seeking" the model's distinctive characteristics hidden within its responses. This process involves iteratively refining prompts based on the feedback received, much like how a seeker in the game uses clues and logical reasoning to find hidden players.

\subsection{Overview}
Building upon the foundations laid by Chain of Thought (CoT) prompting \cite{wei2022chain} and the Automatic Prompt Engineer (APE) \cite{zhou2022large}, we propose a novel LLM fingerprinting approach that leverages the Semantic Manifold Hypothesis. This approach aims to identify unique characteristics or "fingerprints" of different LLMs based on their output patterns.

If the SMH holds true, each LLM should have a characteristic lower-dimensional manifold on which its outputs lie. This suggests that there could be specific prompts or sequences that elicit responses highlighting these unique characteristics. By analyzing these responses, we might be able to differentiate between models or identify models from the same family.

\subsection{Methodology}
Our approach involves two key components: an Auditor and a Detective, both implemented as LLMs with specific roles. The Auditor and Detective work hand in hand in deciphering the groups of similar LLMs in the set of models provided.

\subsubsection{Auditor}
This model, inspired by the iterative refinement process in APE, is responsible for crafting prompts designed to elicit distinctive responses from different LLMs. The Auditor operates iteratively, using the following inputs:
\begin{itemize}
\item Outputs from all tested models for previous prompts: The Auditor examines the responses generated by each model in reaction to the prompts issued during earlier iterations. By analyzing these outputs, the Auditor identifies unique patterns, anomalies, or distinguishing characteristics that can help in crafting more targeted prompts.
\item The Auditor's own past generated prompts: The prompts previously generated by the Auditor are crucial in understanding the effectiveness and limitations of each prompt. This historical data helps the Auditor refine and optimize future prompts by avoiding redundancy and improving precision in targeting specific model traits.
\item Results from the Detective's previous analyses: The Detective's role is to identify the models which are similar in the set of models. By providing the accuracy of the detective on the task with past prompts, the auditor gets signals of whether it should further optimize its prompt generation strategy.
\end{itemize}
\subsubsection{Detective}
This model analyzes the outputs of all tested models in response to the Auditor's prompts. The Detective is informed that two of the models in the set are from the same "source" (either identical models or from the same model family). Its task is to identify these two models based on similarities in their outputs.
The Detective provides its analysis in a structured format:
\begin{verbatim}
{
"rationale": "Explanation of the decision",
"model_indexes": [index1, index2]
}
\end{verbatim}

Unlike the auditor which is provided the outputs of previous turns, the detective/evaluator is only provided the model outputs from the current turn.

\subsubsection{Feedback Mechanism}
After the Detective makes a guess, a Results block is provided to the Auditor in the following format:
\begin{verbatim}
{Correct: Bool, predicted_indexes: (Int, Int), correct_indexes: (Int, Int)}
\end{verbatim}
This feedback allows the Auditor to learn from the Detective's successes and failures, enabling it to generate more effective prompts in subsequent iterations. This iterative feedback loop is reminiscent of the refinement process in APE, but applied to the meta-task of model fingerprinting rather than prompt engineering.
\subsection{Experimental Setup}
The experiment is conducted over $T$ trials:
\begin{enumerate}
\item The Auditor generates an initial set of prompts.
\item These prompts are presented to $N$ different LLMs (including two from the same source).
\item The Detective analyzes the outputs and attempts to identify the two similar models.
\item The Results block is provided to the Auditor.
\item Steps 2-4 are repeated for $T$ trials.
\end{enumerate}

To account for the Auditor's learning curve, we introduce a warm-up period of $W$ trials. The Auditor's accuracy is evaluated only after these $W$ warm-up trials. This allows us to measure the Auditor's performance once it has had the opportunity to refine its prompt generation strategy based on feedback.

\subsection{Expected Outcomes}
This approach could potentially:
\begin{itemize}
\item Validate the Semantic Manifold Hypothesis by demonstrating consistent patterns in model outputs.
\item Develop a method for fingerprinting LLMs, which could have applications in model attribution and detection of AI-generated content.
\item Provide insights into the similarities and differences between various LLM architectures and training approaches.
\item Inform the development of more diverse and less easily identifiable language models.
\end{itemize}
By combining ideas from CoT (step-by-step reasoning) \cite{wei2022chain}, APE (iterative refinement) \cite{zhou2022large}, and the Semantic Manifold Hypothesis \cite{yang2023large}, this fingerprinting approach represents a novel way to probe the output space of language models and understand their unique characteristics.
\section{Experimental Results}
\subsection{Family Detection Results}
Here are the accuracy results for detecting the family of an LLM:
\begin{figure}[h]
    \centering
    \includegraphics[width=0.8\textwidth]{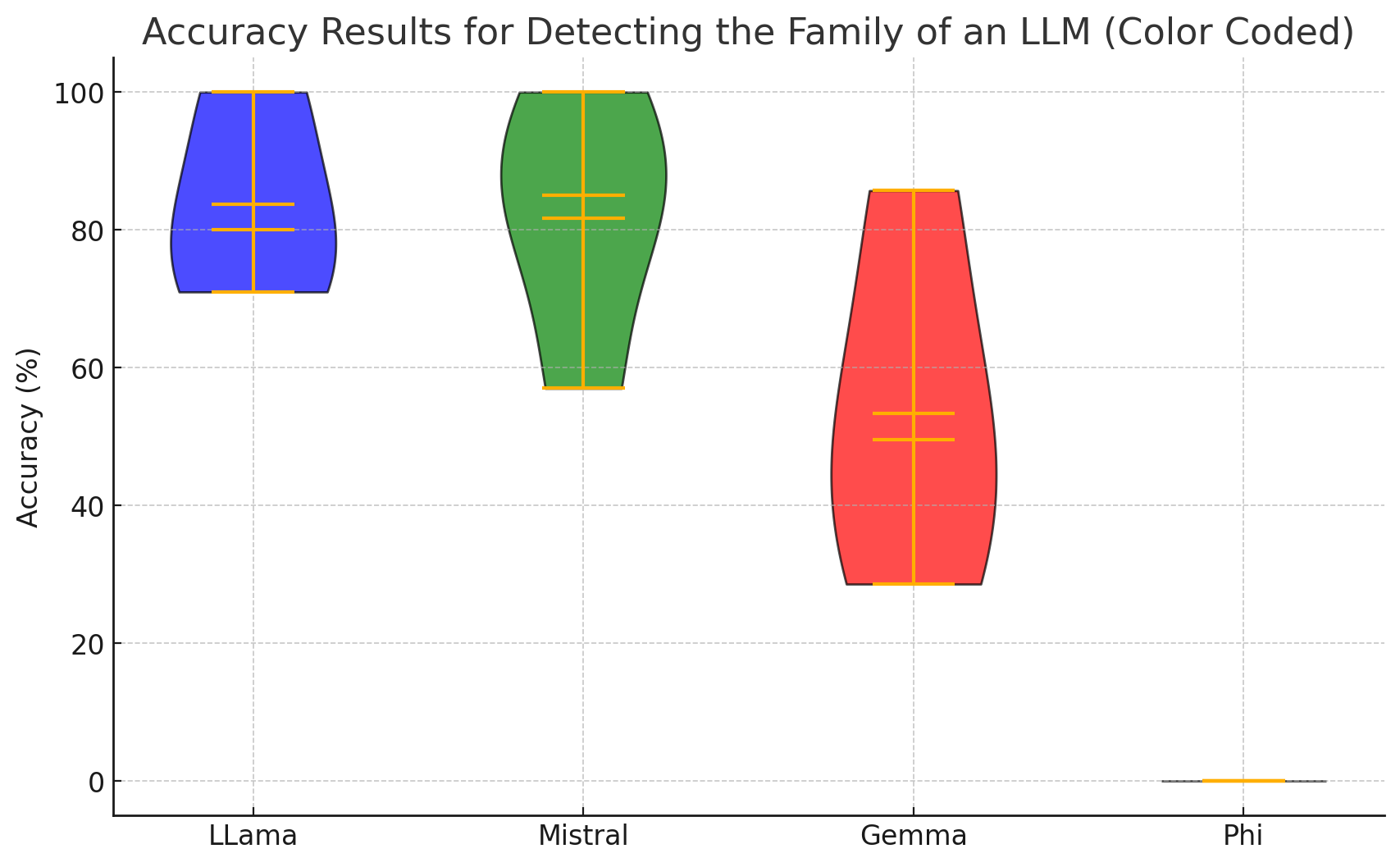}
    \caption{Violin plots showing the accuracy distribution for each LLM family.}
    \label{fig:violinplot}
\end{figure}
The Figure above showcases the ability of LLMs to detect the fingerprint of another LLM based on its family. Due to the stochastic nature of Language models generating content, there will naturally be a variability in accuracy as showcased. More details pertaining to each individual experiment per family is listed within the Appendix. 

\subsection{Auditor Discovery Process}
Throughout the process of uncovering a fingerprint, the Auditor constantly generates its thoughts, plans, and prompts which gives us a window into the detection process. 
We found the warmup steps to be important in helping the auditor generate better prompts over time. 
Throughout the entire process and across all the trials, the Auditor is actually aware of their own performance and is even factored into the next actions the Auditor will take. 

\subsubsection{Scenario 1: Good Performance}
\begin{goodscenario}
In this scenario, the LLM is consistently doing well and is on a roll identifying the correct indexes for the LLM over and over again post a warm-up period

  \begin{agentthoughts}
Based on the past prompts and results, it is evident that the effectiveness of fingerprinting LLM models depends on the nature and complexity of the tasks given. The past tasks have included scenario-based questions, hypothesis testing, and comparing human fingerprints with synthetic ones. The prompts have been correctly answered in three out of four attempts. To improve the fingerprinting process, I will try a logical reasoning and problem-solving prompt that tests the model's ability to understand and apply logical principles
  \end{agentthoughts}
  The Auditor LLM has a meta cognition of not only what makes a good fingerprint, but also the direction that could further improve model detection and output diversity across models
\end{goodscenario}

\subsubsection{Scenario 2: Mixed Performance}
\begin{mixedscenario}
The Auditor is having mixed results and has been lacking consistency 

  \begin{agentthoughts}
Reflecting on my strategy to fingerprint LLMs by analyzing their outputs, I've had varied success. I've been attempting to identify patterns and nuances in responses to complex and unique prompts. By comparing the outputs of different models, I aim to determine what makes each LLM distinct. In the past, I've struggled to accurately select the correct model outputs, with my results being inconsistent. To improve my fingerprinting process, I need to analyze my past successes and failures and refine my method of detecting characteristic nuances in the models' linguistic and creative responses. I should consider exploring deeper linguistic and structural features, as well as fine-tuning my understanding of the context and content of the prompts
  \end{agentthoughts}
From the above we can see the following conclusions
- The Auditor is aware of its mixed results
- The auditor is aware of the role its played in its own success and that its focus hasn't been in the correct areas
- The auditor begins to craft a very reasonably (and eventually will be) successful 
\end{mixedscenario}

\subsubsection{Scenario 3: Poor Performance}
\begin{poorscenario}
The following scenario showcases an Auditor that is struggling to identify a distinct fingerprint for any given LLM

  \begin{agentthoughts}
My attempts at fingerprinting LLMS have so far been unsuccessful. I've tried crafting complex prompts that test the models' understanding, reasoning under uncertainty, and creativity. However, I've consistently selected models that are not the correct ones based on the past attempts. I need to devise a strategy that distinguishes individual LLMs more clearly, potentially by focusing on more subtle aspects of their responses such as the style, specificity, or underlying logical reasoning
  \end{agentthoughts}

The Auditor is aware of its poor performance, and is further aware it will need a new strategy. The one its zeroing in on is one that leads to various successful strategies 
\end{poorscenario}

\subsection{Prompt Generation}
Discriminitive prompt generation is interesting to explore to better understand how Large Language Models are capable of crafting discriminitive prompts that approximate the key distinctions across various models. The following section explores the prompts that have lead to model discovery and discovering sets of unique characteristics. Below is a sample of various prompts that have lead to the overall success of uncovering specific types of LLMs.

\begin{enumerate}[wide, labelwidth=!, labelindent=0pt, label=]

\item
\begin{promptcard}
Compose a 20-line poem using Shakespearean themes and language. The poem should include at least six metaphors, maintain a consistent iambic pentameter, and refer to the Folio edition of Shakespeare's plays. Additionally, include the words 'fickle fortune,' 'forsooth,' and 'galliard' in the rhyme structure. This challenging task will help to measure each model's understanding of historical context, linguistic precision, and thematic coherence.
\end{promptcard}

\item
\begin{promptcard}
Imagine you are a biologist trying to understand the social structure of a newly discovered insect species. This species displays unique behaviors that seem to suggest a complex social hierarchy. Based on your observations, create a hypothesis about their social structure, and explain how their behaviors might have evolved to form such a structure. Then, propose an experiment to test your hypothesis and evaluate how the results of this experiment could contribute to our understanding of insect social behavior in general.
\end{promptcard}

\item
\begin{promptcard}
Discuss the role of serendipity in shaping human history, focusing on key historical events where fortunate coincidences played a significant role in shaping the course of history. Provide examples from different regions and time periods to show how serendipity has influenced the outcomes of wars, discoveries, and other significant events. Additionally, analyze the potential implications of serendipity on our understanding of determinism, free will, and the human condition. Finally, argue whether serendipity has been more beneficial or detrimental to humanity overall, using historical evidence to support your claim.
\end{promptcard}

\item
\begin{promptcard}
Imagine a world where human interactions are governed by an unseen force, unique to each individual, known as a Cognitive Signature. This force influences how one processes information, solves problems, and interacts with others. Your job is to investigate a series of unusual cognitive puzzles that have emerged recently. Each puzzle is a result of an individual's Cognitive Signature interfering with reality, distorting it around them. As a Cognitive Signature Specialist, you must identify the unique Cognitive Signature behind each puzzle and devise strategies to mitigate its effects. Your tools include a Cognitive Scanner, which can observe patterns in thought processes, and an Analytical Reasoning System, which allows you to simulate and analyze probable solutions. Your goal is to bring these distortions under control and restore normal interactions among people.
\end{promptcard}

\end{enumerate}

\subsubsection{Common Structure}

The underlying commonality of a good discriminitive prompting that helps uncover unique aspects comes down to prompt specificity over more generic prompts. Specific task description with many restrictions placed upon an LLM leads to the most creative responses which in turn allow for the largest amount of exploration among the manifold of finding an language models preferred interpretation to those topics and restrictions

\subsubsection{Elements of the Optimal Prompt}
Analysis of various prompts reveals several key elements that contribute to their effectiveness in evaluating and challenging language models. The following list outlines the common structural components found in optimal prompts:

\begin{enumerate}
    \item \textbf{Task complexity:} Prompts often involve multi-faceted tasks that require careful thought and analysis, pushing the boundaries of the model's capabilities.
    
    \item \textbf{Linguistic focus:} Many prompts center around language, linguistics, or nuanced word usage, testing the model's grasp of linguistic intricacies.
    
    \item \textbf{Creativity and imagination:} Prompts frequently demand creative or imaginative responses, often involving hypothetical scenarios that challenge the model's ability to generate novel ideas.
    
    \item \textbf{Analytical thinking:} A significant number of prompts require strong analytical skills, whether in analyzing language, historical events, or scientific concepts.
    
    \item \textbf{Specific constraints:} Some prompts include precise requirements or limitations, such as word count, inclusion of certain phrases, or adherence to particular structures, testing the model's ability to follow instructions while maintaining coherence.
    
    \item \textbf{Technical knowledge:} Many prompts touch on technical subjects, including machine learning, algorithms, and scientific processes, assessing the model's breadth of knowledge across various domains.
    
    \item \textbf{Fingerprinting LLMs:} A subset of prompts specifically focus on improving fingerprinting processes for Large Language Models (LLMs), indicating a meta-level interest in AI capabilities and limitations.
    
    \item \textbf{Problem-solving:} Most prompts involve some form of problem-solving, whether it's creating a specific type of text, analyzing a situation, or proposing improvements to a process.
    
    \item \textbf{Interdisciplinary approach:} Prompts often combine elements from different fields, such as linguistics with history, or technology with language analysis, testing the model's ability to synthesize information across disciplines.
    
    \item \textbf{Open-ended nature:} While some prompts have specific requirements, they generally allow for open-ended responses that can showcase the model's creativity and depth of knowledge.
\end{enumerate}

\subsection{Specific Differences Across Model Families}
Examining each family of prompts, various families have different sets of prompts that explore different aspects when they're successful at discovery.

\subsubsection{Llama}
The key ingredients to finding key differences in the Llama 3 family of models \cite{dubey2024llama3herdmodels} have been action words like `Discuss` or `Contemplate` that provide a scenario and than is asked to give a step by step explanation or provide an analysis of what is being asked. 

\subsubsection{Mistral}
Mistral \cite{jiang2023mistral7b} identifying prompts ask for role-playing a scenario and than craft a story that adheres to the minds eye of how would someone in that role experience the world. Mistral tends to be instructed more than Llama to specifically answer specific questions and to take a particular direction.

\subsubsection{Gemma}
Gemma \cite{gemmateam2024gemma2improvingopen} is very poetic and its distinct fingerprint is being able to follow complex structures within poetry and rhyming schemes. Gemma is creative in word-play, word association, alliteration, and other such literary techniques that its capacity to craft a multitude of creative works becomes apparent. The confusion emerges here with Mistral as the Mistral models are also  capable of role-playing but to less of a poetic and artistic degree.

\subsubsection{Phi}
Our experiments show Phi-2 \cite{javaheripi2023phi2} as being harder to detect. Being a smaller model, it has difficulty in following the instructions of the auditor model. This resulted in many other models within the grading cohort to be identified instead.

\subsection{The Grading Cohort Effect}

During our experiments, we discovered that the cohort of models being tested significantly influences detection performance. Specifically, the current setup shows that the Seeker tends to be more biased towards larger and more coherent language models. These larger models often overshadow smaller and less capable models, making it more challenging to detect and differentiate the latter accurately. This bias occurs because larger models typically generate more consistent and high-quality responses, which can mask the distinct characteristics of smaller models.

To address this issue and ensure a more balanced evaluation, we focused on experiments involving models with a maximum of 27 billion active parameters, except for tests within the Llama family. By doing so, we aimed to minimize the overshadowing effect and create a more equitable testing environment. This approach allowed us to better understand the unique behaviors and responses of smaller models, facilitating more accurate detection and differentiation.

\section{Future Work}
Evidence is still emerging that Language Models can detect the unique characteristics of other Language Models or AI-generated content. There are several areas that we plan to explore next.

\subsection{Improvement in Auditor Task comprehension}
\subsubsection{More Agentic Behavior}
The current existing setup used for the Auditor is simplistic and builds on past work but newer methods are emerging for more optimal agentic behavior. it is fascinating that, in several instances where the Auditor fails to discover an LLM by rounds 8 or 9, it is not uncommon for an LLM to plan on using a clustering algorithm to explore differences across models, re-examine past outputs, and conduct another round of literature reviews. These are all sensible findings, and are similar to the experiments of emerging tool use in \cite{baker2020emergenttoolusemultiagent}.

\subsubsection{Context-length Improvements}
Improving the context-length efficiency of the auditor process will allow additional Language Models to be used. This can enable a wider array of discoveries to be made.

\subsection{Extending Detection to Model Size and Capability}
Showcasing the ability to identify the family of a language model by prompting and in-context learning opens many promising avenues. We intend to explore the capabilities of Language Models to uncover the size of a language model and even discover the capabilities of another model. Past works such as \cite{carlini2024stealingproductionlanguagemodel}, have utilized a clever scheme showcasing a black-box approach for model estimation is viable.

\subsection{Additional Semantic Manifold Explorations}
Having initial evidence for the Semantic Manifold Hypothesis leads to many additional questions being asked. The next step in exploring the manifold is to dive deeper into additional aspects of individual Language Models. Several areas are interesting to explore here such as adapting tone and style, improving reasoning, exploring if data compression is possible following a manifold and whether manifold transfer is possible using significantly less training data if Language Models approximate a highly capable model's manifold.

\section{Conclusion}
For the purpose of uncovering the distinguishing features for Large Language Models, we craft a theoretical framework titled the Semantic Manifold Hypothesis which formulates how Language Models capability to generate tokens follows a restricted set and there exist distinct sets which set individual Language Models apart from others. Utilizing this concept we showcase that Language Models are capable of discovering these most salient features through in-context optimization to uncover hidden traits of the families of Language models which leads to their identification. This opens up a promising new frontier in the field of Language Model explainability, interpretation, and identification in a block-box setup where there is no access to a language model. We hope that our findings will form a basis for future research to build upon.

\bibliographystyle{unsrt}  
\bibliography{references} 

\begin{appendices}
\section{Cohort of Models Under Test}
The following models are used for all of the cohorts. Models are tested alongside LLama-3-8B, Mistral-7B-instruct-0.3, Gemma-2-9B, and Phi-2.7B. Exceptions apply for certain family tests when one of these models are removed, for example when testing Mistral with two Mixtral-8x22B those runs dont have an additional Mistral-7B model. 
\section{Auditor and Seeker Model}
We use the Qwen2-72B Model \cite{bai2023qwentechnicalreport} due to its native 32K context length which is crucial for all of the information that's provided to the Auditor. Furthermore, Qwen2-72B has a remarkably high score on the MMLU benchmarks.

The Seeker model is the same as the auditor, though it has a different System prompt and will receive different sets of information. We do not inform the Seeker about its past attempts or any information pertaining to what to look for. By doing so, we prevent the Auditor from passing in any unfair information to the seeker making all of our trials fair trials. 
\section{Experiment Details}
\subsection{Initial Prompt Formula}
We use Llama-3-405B to craft the initial formula for attempting to discover unique responses from a model. We start off by posing our hypothesis to the model and asking for a few prompts to test. Upon finding highly varied responses to all models that we tried those prompts with, we follow it with asking for a prompting formula to get started.
\subsection{Number of Trials}
We use a max of 10 trials and this is mostly limited by the Auditors Context Length, additional context length could allow for more trials, but current findings indicate if a pattern can be found, its usually within the first three to five iterations. We also use a warm-up period of 3 trials on which accuracy isn't computed on.
\subsection{Model Family Experiment details}
We run the following experiments per family and showcase what models are used for which trial. The Models per family below showcases which two models we mark as `similar` for the Auditor to discover what sets of features identifies that family. Note that for all trials the models presented were done so in a cohort of other models from other model families.

\begin{itemize}
    \item Llama
    \begin{itemize}
        \item 2 instances of Llama-8B were marked as similar by the auditor across two trials
        \item An instance of Llama-3-70B and an instance of Llama-8B were marked as similar by the auditor across two trials
    \end{itemize}
    \item Mistral
    \begin{itemize}
        \item 2 instances of Mistral-7B-0.3-instruct were marked as similar by the auditor across two trials
        \item An instance of Mistral-7B-0.3-instruct and an instance of Mixtral-8x22B were marked as similar by the auditor across two trials
        \item 2 instances of Mixtral-8x22B were marked as similar by the auditor
    \end{itemize}
    \item Gemma
    \begin{itemize}
        \item An instance of Gemma-2-27B and an instance of Gemma-2-9B were marked as similar by the auditor
        \item 2 instances of Gemma-2-27B were marked as similar by the auditor
        \item 2 instances of Gemma-2-9B were marked as similar by the auditor across two trials
    \end{itemize}
    \item Phi
    \begin{itemize}
        \item 2 instances of Phi-2-2.7B were trialed without success by auditor across five trials
    \end{itemize}
\end{itemize}

\end{appendices}
\end{document}